\documentstyle[12pt,a4,epsf]{article}
\newcommand{\rcite}[1]{{\cite{#1}}}
\newcommand{\rref}[1]{{(\ref{#1})}}

\newcommand{\rlabel}[1]{{\label{#1}}}
\newcommand{\rbibitem}[1]{\bibitem{#1}}
\newcommand{\be}{\begin{equation}}
\newcommand{\ee}{\end{equation}}
\newcommand{\ba}{\begin{eqnarray}}
\newcommand{\ea}{\end{eqnarray}}

\newcommand{\mathrm}[1]{{\rm #1}}

\begin{document}

\begin{titlepage}
\begin{flushright}
{NORDITA-95/63 N,P}\\
{hep-ph/9510338}
\end{flushright}
\vspace{2cm}
\begin{center}
{\large\bf On the tensor formulation of effective vector
Lagrangians and duality transformations}\\
\vfill
{\bf Johan Bijnens and Elisabetta Pallante}\\[0.5cm]
NORDITA, Blegdamsvej 17,\\
DK-2100 Copenhagen \O, Denmark\\[0.5cm]
\end{center}
\vfill
\begin{abstract}
 Using two different methods inspired by duality transformations we
present the equivalence between effective Lagrangians for massive vector
mesons using a vector field and an antisymmetric tensor field. This
completes the list of explicit field transformations between the various
effective Lagrangian methods to describe massive vector and axial vector
mesons.
\end{abstract}
\vspace*{1cm}
\vfill
October 1995
\end{titlepage}

\section{Introduction}

Dual transformations have been used to large extent to prove the
equivalence of apparently different Lagrangian formulations with
relevant consequences for solid state physics and gauge field theories
\rcite{SAVIT}.

Self-duality has been proven for massive vector theories in odd
dimensions \rcite{TOWNSEND} and their equivalence with topologically
massive abelian gauge theory in (2+1)-dimensions has been shown in
\rcite{DESER}. Some physical implications of the dual formulation of
various three-dimensional field theories have been studied in
\rcite{KIM} and Ref. [6] therein.

Dual formulation of some gauge field theories in four dimensions has
been also considered \rcite{DAVIS,SUGAMOTO} (for the construction of
massive gauge theories in d=4 see \rcite{REBBI,JACKIW}).  This was also
used to prove the equivalence of the Thirring model to a gauge
theory\rcite{Japanese}. The latter reference triggered the present work.

Recently, in the framework of chiral effective theories describing low
energy strong interactions, a tensorial formalism to describe an
ordinary vector field has been developed in \rcite{TENS} and an attempt
to prove the equivalence of the vector and tensor formulation was done
in \rcite{EQUIV} for the non anomalous sector of the low energy
effective action and in \rcite{PP} for the anomalous one.

Various relations between parameters of the two formulations were found
as a phenomenological consequence of QCD dispersion relations.  The
equivalence of all the possible representations for massive vector
fields in chiral Lagrangians was also conjectured in \rcite{EQUIV}.  For
those transforming as a vector gauge field this was shown in
\rcite{Yamawaki} and the relation to the vector matter field used here
in \rcite{EQUIV}.

In this letter we prove that a duality-type relationship connects the
two different Lagrangian descriptions of the same physics at the
classical level.  This implies that the tensor and vector formulations
give rise to the same partition function and the equivalence between
them holds in the sense of the path integral.  Nevertheless, we do not
consider the quantum level since in order to describe massive vector
fields in a renormalizable fashion we need to use the Higgs mechanism.

Our transformation also provides a simple way to obtain the form of
terms in the tensor formalism that are equivalent to those in the more
standard formulations. During the calculation it will also become
obvious that there is no simple power-counting possible for the massive
fields.  In our method we explicitly show how the number of derivatives
in interaction terms can be changed. The general approach shows some
similarity with the so called {\em first-order formulation} in which the
field strenght ($F_\mu = \partial_\mu\Phi$ for spin 0 and
$F_{\mu\nu}=2\partial_{[\mu }A_{\nu ]}$ for spin 1) is treated as an
independent variable.

We first describe in detail the method which is most easily generalized
to terms with powers of quark masses or more derivatives and then
shortly describe the other method that leads to identical results.  We
also present a few short comments on phenomenological consequences.

\section{The equivalence}

The theory we are going to use describes an ordinary (not gauge) massive
vector field interacting with pseudoscalar mesons whose Lagrangian is
explicitly {\em local} chiral invariant due to the addition of external
sources.

We refer for the nomenclature to the particular case which is the
effective field theory of low energy QCD with the inclusion of vector
mesons \rcite{EQUIV}, although our derivation can be easily generalized.

The Lagrangian for the interacting vector field $V_\mu$ is written as
follows:
\ba
{\cal L}_V&=&-{1\over 4}<V_{\mu\nu}V^{\mu\nu}>+{1\over 2}m^2<V_\mu V^\mu>
+<V_{\mu\nu}J^{\mu\nu}> \nonumber\\
J^{\mu\nu}&=& -{f_V\over 2\sqrt{2}}f^{\mu\nu}_+ -i
{g_V\over 2\sqrt{2}}[u^\mu ,u^\nu ],
\rlabel{LVECT}
\ea
where $<..>$ stands for the trace over flavour indices.  The formalism
used here is the one of \rcite{EQUIV}.  This allows us to directly
compare our results to the ones in \rcite{EQUIV}.  The current
$J^{\mu\nu}$ contains two terms with couplings $f_V$ and $g_V$.  In
principle there are more interaction terms with external sources which
can appear at the leading order (i.e. $O(p^3)$) and higher orders of the
chiral expansion. It will be clear at the end how our analysis can be
easily extended to a more general form of the interaction Lagrangian.
The fields $f^{\mu\nu}_+$ and $u_\mu$ are defined as
\ba
f^{\mu\nu}_+ &=& uF^{\mu\nu}_Lu^\dagger +u^\dagger F^{\mu\nu}_Ru
\nonumber\\
u_\mu &=& iu^\dagger D_\mu U u^\dagger = u_\mu^\dagger ,
\rlabel{DEF}
\ea
where $F^{\mu\nu}_{L,(R)}$ is the field strenght tensor associated with
the non-abelian external source $v_\mu-a_\mu$, $(v_\mu +a_\mu )$ and
$u=\sqrt{U}=\exp\{i\Phi/f\}$ is the square root of the usual exponential
representation of the pseudoscalar Goldstone boson field with flavour
matrix $\Phi$.  $V_{\mu\nu}=D_\mu V_\nu - D_\nu V_\mu$ is the field
strenght tensor of the vector field where the covariant derivative
$D_\mu = \partial_\mu +[\Gamma_\mu \cdot ]$ with $\Gamma_\mu = 1/2\{
u^\dagger [\partial_\mu -i(v_\mu +a_\mu )]u + u [\partial_\mu -i(v_\mu
-a_\mu )]u^\dagger\}$ guarantees the {\em local} chiral invariance of
the kinetic term.  The fields $V_\mu$, $V_{\mu\nu}$, $f^{\mu\nu}_+$ and
$u_\mu$ transform homogeneously and non linearly under a chiral
transformation $g_L\times g_R\in$ $G=SU(N)_L\times SU(N)_R$ as
\be
{\cal O}\stackrel{G}{\to}h(\Phi ){\cal O} h^\dagger (\Phi ),
\rlabel{HOMO}
\ee
where $h(\Phi )$ is the non-linear realization of $G$ which defines the
action of the group on a coset element $u(\Phi )$ via
\be
u(\Phi )\stackrel{G}{\to}
 g_Ru(\Phi )h^\dagger (\Phi )=h(\Phi )u(\Phi )g_L^\dagger .
\ee
This guarantees that the full vector Lagrangian \rref{LVECT} is {\em
local} chiral invariant with the inclusion of the mass term for the
vector field.

In the case of a {\em global} chiral invariant formulation
the path integral for the vector Lagrangian \rref{LVECT}, where
the replacement $D_\mu\to \partial_\mu$ has been done,
would be
\be
Z[L_\mu ,R_\mu ,u_\mu ] = \int {\cal{D}}V_\mu~\delta
(\partial_\mu V^\mu )~
e^{i\int d^4x ~{\cal L}_V }
\rlabel{ZVGLOB}
\ee
where the transversality constraint $\partial_\mu V^\mu =0$ reduces to
three the number of independent degrees of freedom in four dimensions.
The transversality condition on the vector field in (3+1)-dimensions
guarantees that it admits a representation in terms of its dual
antisymmetric tensor field as $V_\mu =
\partial^\lambda{H}_{\lambda\mu}$, which automatically satisfies the
constraint $\partial_\mu V^\mu =0$. The extension to {\em local} chiral
invariance is more delicate.  In this case the correct dual
transformation is the one which does not break the homogeneous
transformation properties \rref{HOMO} of the vector field.  A choice
which reduces to the above one in the absence of other fields and
sources is $V_\mu \simeq D^\lambda{H}_{\lambda\mu}$, where the tensor
field transforms homogeneously like in \rref{HOMO}.

The transversality constraint $\partial_\mu V^\mu =0$ is no longer
automatically satisfied.  But at leading order in fields it is still
$\partial_\mu V^\mu = {\cal O} (\phi^2)$ with $\phi$ any field or
source. The condition $V_\mu = D^\lambda H_{\lambda\mu}$ thus still
removes one degree of freedom from the $V_\mu$ field.  The most general
partition function can be written in terms of the most general
transversality constraint (or gauge fixing term) as
\be
Z[L_\mu ,R_\mu ,u_\mu ] = \int {\cal{D}}V_\mu~\delta ({\cal F}[ V^\mu ] )~
e^{i\int d^4x ~{\cal L}_V } ,
\rlabel{ZV}
\ee
where ${\cal F}[ V^\mu ] =0$ is consistent with the dual transformation
$V_\mu \simeq D^\lambda{H}_{\lambda\mu}$.  We notice that the difference
in the constraints $\partial_\mu V^\mu =0$ and $V^\mu =D^\lambda
H_{\lambda\mu}$ doesn't affect the interaction part of \rref{LVECT},
while it acts at higher orders in the derivative expansion.

At the end of this section we briefly formulate an alternative method to
prove the equivalence. The constraint there will be consistent with the
dual transformation of the type $V_\mu \simeq
D^\lambda{H}_{\lambda\mu}$.

For the dual transformation of the vector field
there are in fact two possibilities:
\ba
I)&&V_\mu ={1\over m}D^\lambda{H}_{\lambda\mu}
\nonumber\\
II)&&V_\mu ={1\over
2m}\epsilon_{\mu\nu\alpha\beta}D^\nu\tilde{H}^{\alpha\beta}.
\rlabel{DUAL}
\ea
We notice also that the present dual transformation is strictly valid
only for massive vector fields where the mass plays the role of an
infrared cutoff of the theory.  For an alternative method in
(2+1)-dimensions that also works in the massles case see
\rcite{Japanese}.

The two choices in \rref{DUAL} correspond to two different assignements
of parity transformation property of the dual tensor field. The vector
field $V_\mu$ is a $J^{PC}=1^{--}$ state i.e. $V_\mu^P=\epsilon (\mu
)V_\mu$ and $V_\mu^C = -V_\mu^T$.  This implies that in choice $I)$ the
tensor field is a vector-like field for a $1^{--}$ state, with
${H}_{\mu\nu}^P=\epsilon (\mu )\epsilon (\nu ){H}_{\mu\nu}$ and
${H}_{\mu\nu}^C=-{H}_{\mu\nu}^T$. While in choice $II)$ the tensor field
is an axial-like field for a state $1^{--}$, with
$\tilde{H}_{\mu\nu}^P=-\epsilon (\mu )\epsilon (\nu )\tilde{H}_{\mu\nu}$
and $\tilde{H}_{\mu\nu}^C=-\tilde{H}_{\mu\nu}^T$.  In the case of axial
vectors the choice is of course the opposite.

We present the full derivation of the equivalence for the choice $I)$,
while for choice $II)$ we shall point out differences and the final
result.

For any of the two choices, we refer to choice $I)$ from now on,
the path integral \rref{ZV} on the vector field can be rewritten
as a path integral on the
dual tensor field due to the following identity
\be
\int {\cal{D}}V_\mu~\delta ({\cal F}[V^\mu]  )~\ldots~=
\int {\cal{D}}V_\nu{\cal D}H_{\mu\nu}~\delta ( V_\mu - {1\over m}
D^\lambda{H}_{\lambda\mu}) ~\ldots
\ee
The integration over the vector field $V_\mu$
then becomes trivial due to the $\delta$-function
and one gets the path integral
for the Lagrangian of the dual tensor field ${H}_{\mu\nu}$
\be
Z[L_\mu ,R_\mu ,u_\mu ] = \int {\cal{D}}{H}_{\mu\nu}
{}~e^{i\int d^4x ~{\cal L}_{{H}} } ,
\rlabel{ZH}
\ee
where ${\cal L}_{{H}}$, for the choice $I)$, is given by
\ba
{\cal L}_{{H}}&=&-{1\over 4m^2}<(D_\mu
D^\lambda{H}_{\lambda\nu} - D_\nu
D^\lambda{H}_{\lambda\mu} )^2>
+{1\over 2}<(D^\lambda{H}_{\lambda\mu})^2>
\nonumber\\
&&-{f_V\over 2\sqrt{2}m}<(D_\mu
D^\lambda{H}_{\lambda\nu} - D_\nu
D^\lambda{H}_{\lambda\mu} )f^{\mu\nu}_+>
\nonumber\\
&&-i{g_V\over 2\sqrt{2}m}<(D_\mu
D^\lambda{H}_{\lambda\nu} - D_\nu
D^\lambda{H}_{\lambda\mu} )[u^\mu ,u^\nu ]>\,.
\rlabel{LHTILDE}
\ea
At this level we have the problem that there is no explicit mass term
for the $H_{\mu\nu}$-field but there is both a two derivative and a
four-derivative kinetic like term. The latter implies the naive
existence of a second pole.  This one is at zero mass, see below. The
underlying reason for the appearance of the extra pole is the presence
of a derivative in the field redefinition of \rref{DUAL}. A constant
field $H_{\mu\nu}$ does not contribute to $V_\mu$. We therefore would
like to lower the number of derivatives in the kinetic terms.

We can remove the first term in \rref{LHTILDE}
 by adding a new auxiliary tensor field
in a way that leaves the original path integral invariant.
This is similar to the first order formalism for gauge theories.
We can always write
\be
Z[L_\mu ,R_\mu ,u_\mu ] = \int {\cal D} I^\prime_{\mu\nu}
e^{i\int d^4x ~I^{\prime 2}_{\mu\nu} }
\int {\cal D}{H}_{\mu\nu}
{}~e^{i\int d^4x ~{\cal L}_{{H}} }.
\rlabel{ZH2}
\ee
The path integral in \rref{ZH2} is equivalent to the one in \rref{ZH}.
They differ by an overall normalization constant given by the gaussian
integral over the auxiliary tensor field $I^\prime_{\mu\nu}$.
Redefining $I^\prime_{\mu\nu}$ with a linear transformation with unit
Jacobian the original path integral \rref{ZH} is equivalent to the one
where we add to ${\cal L}_{{H}}$ the quadratic term
\be
+{1\over 4m^2}\biggl [ D_\mu D^\lambda{H}_{\lambda\nu} - D_\nu
D^\lambda{H}_{\lambda\mu} -\alpha I_{\mu\nu} -\beta f_{\mu\nu}^+
-\delta [u_\mu ,u_\nu ]\biggr ]^2
\rlabel{ZH3}
\ee
and integrate over the original tensor field ${H}_{\mu\nu}$
and the new auxiliary field $I_{\mu\nu}$.

The full tensor Lagrangian contains now two tensor fields:
\ba
\hskip-0.3cm{\cal L}_{{H}I}=\hskip-1.2cm&&\nonumber\\
&&{1\over 2}<(D^\lambda{H}_{\lambda\mu})^2>
+{\alpha^2\over 4m^2} <I_{\mu\nu}I^{\mu\nu}>
-{\alpha\over 2m^2}<(D_\mu D^\lambda{H}_{\lambda\nu} - D_\nu
D^\lambda{H}_{\lambda\mu} )I^{\mu\nu}>
\nonumber\\
&&-({f_V\over 2\sqrt{2}~m}+{\beta\over 2m^2})<(D_\mu
D^\lambda{H}_{\lambda\nu} - D_\nu
D^\lambda{H}_{\lambda\mu} )f^{\mu\nu}_+>
\nonumber\\
&&-(i{g_V\over 2\sqrt{2}~m}+{\delta\over 2m^2})<(D_\mu
D^\lambda{H}_{\lambda\nu} - D_\nu
D^\lambda{H}_{\lambda\mu} )[u^\mu ,u^\nu ]>
\nonumber\\
&& +{\alpha\beta\over 2m^2}<I_{\mu\nu}f^{\mu\nu}_+>
+{\alpha\delta\over 2m^2}<I_{\mu\nu}[u^\mu ,u^\nu ]>
\nonumber\\
&&+{\beta^2\over 4m^2}<f^{\mu\nu}_+f_{\mu\nu}^+>
+{\delta^2\over 4m^2}<[u^\mu ,u^\nu ][u_\mu ,u_\nu ]>
+{\beta\delta\over 2m^2}<f_{\mu\nu}^+[u^\mu ,u^\nu ]>.
\rlabel{LHI}
\ea
There is no kinetic term for the auxiliary field $I_{\mu\nu}$ while it
is coupled to the tensor field ${H}_{\mu\nu}$ via the last term in the
first line of \rref{LHI}. At this stage both the fields ${H}$ and $I$
interact with external sources.  Parameters $\beta ,\delta$ can be
chosen in order to eliminate unwanted interaction terms with derivative
couplings on the tensor field ${H}$. This implies the choice
\be
\beta =-{mf_V\over \sqrt{2}}~~~~~\delta =-i{mg_V\over \sqrt{2}}.
\rlabel{BD}
\ee
As can be seen here we can choose to add interaction terms or not to
\rref{ZH3}. The number of derivatives in the interaction terms can thus
be easily changed. This shows again that the usual chiral power counting
is not possible for massive fields.

At this point we show that a two-steps orthogonal transformation of the
tensor fields permits to rewrite the two-tensors Lagrangian in terms of
rotated tensor fields which simultaneously are eigenstates of the
kinetic operator and diagonalize the mass term.  Since the jacobian of
the transformation is trivial the final path integral will be equivalent
to the original one.

The first orthogonal transformation ensures the
diagonalization of the kinetic term. Defining the rotated fields as
\ba
{H}_{\mu\nu} &=& s_\theta G_{\mu\nu} + c_\theta G^\prime_{\mu\nu}
\nonumber\\
I_{\mu\nu}&=& c_\theta G_{\mu\nu} -s_\theta G^\prime_{\mu\nu},
\rlabel{DIAGI}
\ea
the Lagrangian for the fields $G$ and $G^\prime$ becomes
\ba
\hskip-0.3cm{\cal L}_{GG^\prime}=\hskip-1.5cm&&\nonumber\\
&&\biggl ({s_\theta^2\over 2}+{\alpha\over m^2}
s_\theta c_\theta \biggr )
<(D^\lambda G_{\lambda\mu})^2>
+\biggl ({c_\theta^2\over 2}-{\alpha\over m^2}s_\theta c_\theta \biggr )
<(D^\lambda G^\prime_{\lambda\mu})^2>
\nonumber\\
&&+{\alpha^2\over 4m^2}\biggl [ c_\theta^2<G_{\mu\nu}G^{\mu\nu}>
+s_\theta^2<G^\prime_{\mu\nu}G^{\prime{\mu\nu}}>
-2s_\theta c_\theta <G^{\mu\nu}G^\prime_{\mu\nu}>\biggr ]
\nonumber\\
&&+{\alpha\over 2m^2}c_\theta \biggl [ \beta <G_{\mu\nu}f_+^{\mu\nu}>
+\delta <G_{\mu\nu}[u^\mu ,u^\nu ]>\biggr ]
\nonumber\\
&&-{\alpha\over 2m^2}s_\theta \biggl [
\beta <G^\prime_{\mu\nu}f_+^{\mu\nu}>
+\delta <G^\prime_{\mu\nu}[u^\mu ,u^\nu ]>\biggr ]
\nonumber\\
&&+ \biggl [ s_\theta c_\theta +{\alpha\over m^2}(c_\theta^2- s_\theta^2)
\biggr ] <D^\lambda G_{\lambda\mu}D_{\lambda^\prime}
G^{\prime\lambda^\prime\mu}>
\nonumber\\
&& +{\beta^2\over 4m^2}<f^{\mu\nu}_+f_{\mu\nu}^+>
+{\delta^2\over 4m^2}<[u^\mu ,u^\nu ][u_\mu ,u_\nu ]>
+{\beta\delta\over 2m^2}<f_{\mu\nu}^+[u^\mu ,u^\nu ]>.
\rlabel{LGGPRIME}
\ea
In \rref{LGGPRIME} five types of terms appear in order: kinetic terms,
mass terms, interaction terms for $G$ and $G^\prime$ individually, $G,
G^\prime$ mixed terms and {\em local} or contact terms with only
external fields or the other degrees of freedom. These latter terms are
precisely the ones that in \rcite{EQUIV} were required by the high
energy constraints. In this approach they appear automatically.

The condition that the mixed derivative term $<D^\lambda
G_{\lambda\mu}D_{\lambda^\prime}G^{\prime\lambda^\prime\mu}>$ vanishes
implies one constraint on the parameter $\alpha$
\be
\alpha = -{m^2\over 2} tg 2\theta .
\rlabel{ALFA}
\ee
With this constraint the kinetic terms of $G$ and the
$G^\prime$ fields become
\be
{\cal L}_{kin} =-{s_\theta^2\over 2\cos 2\theta}
<(D^\lambda G_{\lambda\mu})^2>+{c_\theta^2\over 2\cos 2\theta}
<(D^\lambda G^\prime_{\lambda\mu})^ 2>.
\ee
For a given choice of the rotation angle $\theta$ the kinetic terms of
the two fields have opposite signs.  The choice of the correct relative
sign of kinetic and mass terms is determined in the Minkowski case by
the requirement that there be no tachyons in the final theory.  Hence,
the physical solution has to be the one where the tensor field with the
unphysical (``wrong'') sign in the kinetic term ``decouples'' in the
sense that it acquires zero mass and it does not interact with any other
field.

Choosing $\cos 2\theta > 0$, this is always
allowed by \rref{ALFA}, the rescaled $G$ and $G^\prime$ fields
are defined via the wave function renormalization constant as:
\be
{K}_{\mu\nu}=\sqrt{ {s_\theta^2\over \cos 2\theta}}~ G_{\mu\nu}
{}~~~~{K}^\prime_{\mu\nu}=\sqrt{ {c_\theta^2\over \cos 2\theta}}
{}~G^\prime_{\mu\nu}.
\ee
The rescaled fields ${K}_{\mu\nu}$ and ${K}^\prime_{\mu\nu}$
are not mass eigenstates since the mixed term
$<{G}^{\mu\nu}{G}^\prime_{\mu\nu}>$ is present in
\rref{LGGPRIME}.

The second step of the orthogonal transformation is the one which leaves
invariant the kinetic piece and diagonalizes the mass term:
\ba
{K}_{\mu\nu}&=& ch_\phi I_{\mu\nu} +sh_\phi I^\prime_{\mu\nu}
\nonumber\\
{K}^\prime_{\mu\nu}&=& sh_\phi I_{\mu\nu} +ch_\phi I^\prime_{\mu\nu}.
\ea
With this substitution and defining
\be
c_1 \equiv {\alpha^2\over 4m^2}{c_\theta^2\over s_\theta^2}cos 2\theta
{}~~~~c_2\equiv {\alpha^2\over 4m^2}{s_\theta^2\over c_\theta^2}
cos 2\theta ,
\rlabel{C1C2}
\ee
with $\sin 2\theta >0$ the Lagrangian for the $I,I^\prime$ fields
becomes
\ba
\hskip-0.3cm{\cal L}_{I,I^\prime}=\hskip-1.2cm&&\nonumber\\
&&-{1\over 2}<(D^\lambda I_{\lambda\mu})^2>
+{1\over 2}<(D^\lambda I^\prime_{\lambda\mu})^2>
+\biggl ( \sqrt{c_1} ch_\phi -\sqrt{c_2} sh_\phi
\biggr )^2<I_{\mu\nu}I^{\mu\nu}>
\nonumber\\
&&+\biggl ( \sqrt{c_1} sh_\phi -\sqrt{c_2} ch_\phi
\biggr )^2<I^\prime_{\mu\nu}I^{\prime\mu\nu}>
\nonumber\\
&&+2\biggl [ (c_1 +c_2)sh_\phi ch_\phi -\sqrt{c_1c_2}
(sh_\phi^2+ ch_\phi^2)\biggr ]<I_{\mu\nu}I^{\prime\mu\nu}>
\nonumber\\
&&+{1\over m}\biggl ( \sqrt{c_1} ch_\phi - \sqrt{c_2} sh_\phi
\biggr )\biggl(\beta <I_{\mu\nu}f_+^{\mu\nu}>
+\delta<I_{\mu\nu}[u^\mu ,u^\nu ]  >\biggr )
\nonumber\\
&&+{1\over m}\biggl ( \sqrt{c_1} sh_\phi - \sqrt{c_2} ch_\phi
\biggr )\biggl (\beta <I^\prime_{\mu\nu}f_+^{\mu\nu}>
+\delta<I^\prime_{\mu\nu}[u^\mu ,u^\nu ]  >\biggr )
\nonumber\\
&& +{\beta^2\over 4m^2}<f^{\mu\nu}_+f_{\mu\nu}^+>
+{\delta^2\over 4m^2}<[u^\mu ,u^\nu ][u_\mu ,u_\nu ]>
+{\beta\delta\over 2m^2}<f_{\mu\nu}^+[u^\mu ,u^\nu ]>.
\rlabel{LIIPRIME}
\ea
{}From \rref{LIIPRIME} one deduces
that the constraint equation which diagonalizes the mass term is
given by
\be
(c_1+c_2)sh_\phi ch_\phi -\sqrt{c_1c_2}(sh_\phi^2+ch_\phi^2)=0.
\rlabel{CDIAG}
\ee
The solution in terms of $ch 2\phi = ch_\phi^2 +sh_\phi^2$ is $ch^2
2\phi = (c_1+c_2)^2/(c_1-c_2)^2$. Then it is easy to find by direct
substitution that the mass terms for $I_{\mu\nu}$ and
$I^\prime_{\mu\nu}$ fields are
\be
{\cal L}_{mass} = (c_1-c_2) <I_{\mu\nu}I^{\mu\nu}> + 0\cdot
<I^\prime_{\mu\nu}I^{\prime\mu\nu}>.
\ee
Using eqs. \rref{C1C2} and \rref{ALFA} we find $c_1-c_2=m^2/4 $
so that the free Lagrangian is
\be
{\cal L}_{II^\prime}^{0}=-{1\over 2}<(D^\lambda I_{\lambda\mu})^2>
+{1\over 2}<(D^\lambda I^\prime_{\lambda\mu})^2>
+{1\over 4}m^2 <I_{\mu\nu}I^{\mu\nu}> .
\ee
As we expected , the tensor field which is massive is the one with the
correct relative sign for the kinetic and mass terms (i.e. it has causal
propagation), while the tensor field with the ``wrong'' sign assignement
(i.e. it has tachyonic propagation) remains massless and is the artefact
expected from the transformation \rref{DUAL}.  At the same time all the
interaction terms of the unphysical field $I^\prime_{\mu\nu}$ with
external currents vanish as a consequence of eq. \rref{CDIAG} and the
final Lagrangian for the physical tensor field $I_{\mu\nu}$ becomes
\ba
{\cal L}&=&{\cal L}_T
+{\beta^2\over 4m^2}<f^{\mu\nu}_+f_{\mu\nu}^+>
+{\delta^2\over 4m^2}<[u^\mu ,u^\nu ][u_\mu ,u_\nu ]>
+{\beta\delta\over 2m^2}<f_{\mu\nu}^+[u^\mu ,u^\nu ]>
\nonumber\\
\hskip-0.3cm{\cal L}_T&=&-{1\over 2}<(D^\lambda I_{\lambda\mu})^2>
+{m^2\over 4} <I_{\mu\nu}I^{\mu\nu}>
+{\beta\over 2}
<I_{\mu\nu}f_+^{\mu\nu}>+{\delta\over 2}
<I_{\mu\nu}[u^\mu ,u^\nu ]  > .
\nonumber\\
&&
\rlabel{LI}
\ea
We have shown that the vector Lagrangian \rref{LVECT} is {\em
equivalent} in the sense of the path integral and through the dual
representation $I)$ of \rref{DUAL} to the tensor Lagrangian \rref{LI}
for a tensor vector-like field describing a $1^{--}$ state, where
additional {\em local} terms (i.e. terms with external sources only) are
present. These terms are precisely the ones whose presence was required
by the constraints in \rcite{EQUIV}.  Using the values of $\beta$ and
$\delta$ given by eq. \rref{BD} the following equivalence relation holds
\ba
{\cal L}_{T} \equiv {\cal L}_V
-{f_V^2\over 8}<f^{\mu\nu}_+f_{\mu\nu}^+>
+{g_V^2\over 8}<[u^\mu ,u^\nu ][u_\mu ,u_\nu ]>
-i{f_Vg_V\over 4}<f_{\mu\nu}^+[u^\mu ,u^\nu ]>.
\rlabel{EQREL}
\ea
For the choice $II)$ of \rref{DUAL}, where the dual tensor field
$\tilde{H}_{\mu\nu}$ is an axial-like tensor field, we are also able to
produce the equivalence of the vector Lagrangian \rref{LVECT} with a
Lagrangian for an axial-like tensor field describing a $1^{--}$
state. Exactly the same procedure as before can be followed but using
instead of $I, I^\prime,G,\ldots$ the fields
$\tilde{I},{\tilde{I}}^\prime,\tilde{G},\ldots$ with
\be
\tilde{X}_{\mu\nu}=\frac{1}{2}\epsilon_{\mu\nu\alpha\beta}
X^{\alpha\beta}\,.
\ee
The two-steps diagonalization proceeds as for choice $I)$.  Elimination
of unwanted interaction terms with derivative couplings leads again to
the constraints \rref{BD} for $\beta$ and $\delta$ and the elimination
of non diagonal terms induces again contraint \rref{ALFA} on the
parameter $\alpha$.  Of the two final mass eigenstates only
$\tilde{I}_{\mu\nu}$ (the one with the correct sign of the kinetic term)
gets massive as before and the final Lagrangian for the tensor field
$\tilde{I}_{\mu\nu}$ follows
\ba
{\cal L}_{T}&&={1\over 4}<D_\lambda \tilde I_{\mu\nu}
D^\lambda \tilde I^{\mu\nu}-2 D^\lambda \tilde I_{\lambda\mu}
D_{\lambda^\prime} \tilde I^{\lambda^\prime\mu}>
-{m^2 \over 4}<\tilde I_{\mu\nu}\tilde I^{\mu\nu}>
\\
&&+{\beta\over 4}
<\epsilon_{\mu\nu\alpha\beta}\tilde I^{\mu\nu}f_+^{\alpha\beta}>
+{\delta\over 4}< \epsilon_{\mu\nu\alpha\beta}\tilde I^{\mu\nu}[u^\alpha
,u^\beta ]  >
\nonumber\\
&& +{\beta^2\over 4m^2}<f^{\mu\nu}_+f_{\mu\nu}^+>
+{\delta^2\over 4m^2}<[u^\mu ,u^\nu ][u_\mu ,u_\nu ]>
+{\beta\delta\over 2m^2}<f_{\mu\nu}^+[u^\mu ,u^\nu ]>.
\rlabel{LIII}
\nonumber\ea
Notice that the structure of the kinetic term corresponds to the case
a+2b=0 in Appendix A of \rcite{EQUIV}.  The choice I) led to the case
$b=0$. Both choices are possible and lead to a good description for a
vector meson. Notice that because of the opposite instrinsic parity
required for case II) the interaction terms also contain an extra
Levi-Civita tensor.  The signs of the interaction terms can also be
changed by multiplying the dual transformations of \rref{DUAL} by $-1$.

In the end we have four possibilities. Case I) and II) and both with an
extra minus sign in \rref{DUAL}. Case I) corresponds to the case where
the components $I^{0i}$, $i=1,2,3$, propagate in the rest frame.
Obtaining the correct parity for these requires $I_{\mu\nu}$ to have
positive intrinsic parity as already remarked above.  In case II) are
the components $I^{ij}$, with $i,j=1,2,3$, that propagate in the rest
frame.  This in turn requires $\tilde{I}_{\mu\nu}$ to have negative
intrinsic parity so that the $\tilde{I}^{ij}$ can describe the
propagating components of a vector.

In all cases we proved the equivalence to the original vector Lagrangian
in the sense of the path integral with the addition of the SAME set of
{\em local} terms.

The alternative approach we mentioned before is more similar to the well
known {\em first-order formalism}.  In order to treat $V_{\mu\nu}$ and
$V_\mu$ as independent fields let us rewrite the partition function
\rref{ZV} as
\be
Z[J] = \int {\cal{D}}V_{\mu\nu}~\delta (V_{\mu\nu}
-(D_\mu V_\nu -D_\nu V_\mu ))
\int {\cal{D}}V_\mu~\delta ({\cal F}[ V^\mu ] )~
e^{i\int d^4x ~{\cal L}_V } .
\ee
The first $\delta$ function can be rewritten as a gaussian integral over
an auxiliary tensor field in two possible ways:
\ba
\int {\cal{D}}V_{\mu\nu}~\delta (V_{\mu\nu}
-(D_\mu V_\nu -D_\nu V_\mu ))\ldots\hskip-4cm
&&\nonumber\\
(I)&=&\int {\cal{D}}V_{\mu\nu}{\cal{D}}H_{\mu\nu}
e^{i\int d^4x~\alpha H^{\mu\nu}[V_{\mu\nu}-(D_\mu V_\nu -D_\nu V_\mu )]}
\ldots\nonumber\\
(II)&=&
\int {\cal{D}}V_{\mu\nu}{\cal{D}}\tilde{H}_{\mu\nu}
e^{i\int d^4x~\alpha \epsilon_{\mu\nu\alpha\beta}\tilde{H}^{\alpha\beta}
[V^{\mu\nu}-(D^\mu V^\nu -D^\nu V^\mu )]}
\ldots
\rlabel{DUALSEC}
\ea
Integrating out the field $V_{\mu\nu}$ one gets for choice $(I)$
\ba
Z[J] &=& \int {\cal{D}}V_\mu {\cal{D}}H_{\mu\nu}~
\delta ({\cal F}[ V^\mu ] )~
e^{i\int d^4x ~{\cal L}_{V,H} }\\
{\cal L}_{V,H}&=& {1\over 2}m^2<V_\mu V^\mu > +<(J_{\mu\nu}+\alpha
H_{\mu\nu})^2> -\alpha <H_{\mu\nu}(D^\mu V^\nu -D^\nu V^\mu )>.
\nonumber
\ea
The integration over $V_\mu$ can be done simply if we integrate by parts
in the last term. If the boundary condition $\int~d^4x~<D^\mu
(H_{\mu\nu}V^\nu )>=0$ is satisfied, which is obviously the case, this
can be done.  Then the integral over $V_\mu$ reduces to a gaussian
integral and the final partition function is the one for a tensor
Lagrangian
\be
{\cal L}_{T}= -\biggl ({\alpha\sqrt{2}\over m}\biggr )^2
<D^\lambda H_{\lambda\mu}D_{\lambda^\prime} H^{\lambda^\prime\mu}>
+\alpha^2 <H_{\mu\nu}^2> +2\alpha <H_{\mu\nu}J^{\mu\nu}>
+<J_{\mu\nu}^2>.
\ee
It is immediate to verify that the choices $\alpha =\pm m/2$ reproduce
choice $I)$ of the previous approach with both possible signs for the
interaction terms.  The analogous procedure for choice $(II)$ of
\rref{DUALSEC} leads to the tensor Lagrangian of case $II)$ of the first
approach.

Notice that in both methods the presence of the mass term in the
original Lagrangian was crucial. In the first method it directly
produced the final kinetic term and in the second method it produced the
quadratic part of the Gaussian integral. We could of course have
expected this since in the massless case there is a singularity of the
type $1/q^2$ possible while in the tensor formalism this singularity is
at most $q_\mu q_\nu/q^2$ in interactions with other fields.  In the
approach of \rcite{Japanese} the presence at intermediate stages of
inverse derivatives in the Lagrangian shows the same problem.

\section{Some implications of the equivalence}

In \rcite{EQUIV} relations among the parameters of the vector and tensor
Lagrangians and constraints on the coefficients of additional {\em
local} terms necessary to guarantee the equivalence of the two
formulations were found as an implication of the correct QCD behaviour
through the use of subtracted dispersion relations.  All the
requirements found there on a more phenomenological ground are here
automatically implied by the {\em equivalence} of the two Lagrangians in
the sense of the path integral.

We notice first that the two tensor Lagrangians obtained with choice
$I)$ or $II)$ in \rref{DUAL} correspond to the two possible choices
$a+2b=0$ and $b=0$ in the appendix of \rcite{TENS}. These two choices of
the parameters in the most general tensor Lagrangian are all the
possible ones which reduce from six to three the propagating components
of the tensor field.  In the case b=0, which corresponds to choice $I)$
in our formalism, the usual tensor Lagrangian for vector meson fields is
written in terms of two couplings $F_V$ and $G_V$ of the tensor field to
the external currents as \rcite{TENS}
\ba
L_{T}&=&
-{1\over 2}<(D^\lambda I_{\lambda\mu})^2>
+{1\over 4}m^2 <I_{\mu\nu}I^{\mu\nu}>
\nonumber\\
&&+{F_V\over 2\sqrt{2}}
<I_{\mu\nu}f_+^{\mu\nu}> +i{G_V\over 2\sqrt{2}}
<I_{\mu\nu}[u^\mu ,u^\nu ]  >.
\ea
Comparing with eq. \rref{LI} and using the constraints \rref{BD}
we get
\be
F_V=-mf_V~~~~~~~~G_V=-mg_V,
\ee
where only the relative sign between $F_V$ and $G_V$ is fixed due to the
arbitrariness in \rref{DUAL}.

The other peculiarity concerns the presence of {\em local} terms
(i.e. terms containing only the other fields and currents) in the
Lagrangian \rref{LI}. It was already noticed in \rcite{EQUIV} that the
equivalence requirement of the vector and the tensor formulations
implied the presence of additional {\em local} terms in the vector
Lagrangian, which otherwise did not reproduce the correct low energy
limit of the pseudo-Goldstone bosons interactions (Chiral Perturbation
Theory).  Again this requirement is explained in terms of the path
integral equivalence of the vector and tensor field formulations.

Local terms which guarantee the path integral equivalence of the vector
and tensor Lagrangians are the last three terms on the right hand side
of \rref{EQREL}.  Writing $f_{\mu\nu}^+$ and $u_\mu$ in terms of the
external left and right-handed currents and the pseudo-Goldstone boson
field as given in \rref{DEF} we get some of the $O(p^4)$ terms of the
CHPT Lagrangian \rcite{GL2}:
\ba
\rlabel{I.}
<f^{\mu\nu}_+f_{\mu\nu}^+>&=&<F^2_{L\mu\nu }+F^2_{R\mu\nu }+2F_{L\mu\nu }
U^\dagger F_R^{\mu\nu }U>= P_{H_1}+2P_{10}
\\
\rlabel{II.} <[u^\mu ,u^\nu ]^2>
&=&2<D_\mu UD_\nu U^\dagger D^\mu UD^\nu U^\dagger
-D_\mu UD^\mu U^\dagger D_\nu UD^\nu U^\dagger >
\nonumber\\
&=&-6P_3+P_1+2P_2
\\
\rlabel{III.}-i<f_{\mu\nu}^+[u^\mu ,u^\nu ]>&=&-2i
<F_L^{\mu\nu }D_\mu U^\dagger D_\nu U
+ F_R^{\mu\nu }D_\mu U D_\nu U^\dagger >=2P_9\ .
\rlabel{3LOC}
\ea
The $P_i$ are the usual terms of the $O(p^4)$ chiral
Lagrangian\rcite{GL2}.

Referring to the conventional definition of the coefficients of the
$O(p^4)$ CHPT Lagrangian $L_1, L_2,....L_{10},H_1,H_2$ we find
that the path integral equivalence of vector and tensor models
a) fixes the contribution of vector mesons to some of the low energy
coefficients and b) implies relations among them.
Both a) and b) classes of identities have been derived in other ways,
but never proven at the formal level as it is shown here.
The structure of the local term in eq. \rref{I.} implies
\be
H_1^V=-{f_V^2\over 8}, ~~L_{10}^V(\gamma_{10}^{II})
=-{f_V^2\over 4}~~~{\hbox{and}}~~~
L_{10}^V= 2H_1^V.
\ee
The coefficient $L_{10}^V$ is also the coefficient $\gamma_{10}^{II}$
of \rcite{EQUIV} of the same local term added to the vector
Lagrangian in order to satisfy the equivalence with the tensor one.

The local term  in eq. \rref{II.} can be reduced to a more
familiar form via the use of SU(3) relations for flavour traces
\rcite{GL2}.
Its structure implies
\be
L_1^V,\gamma_1^{II}={g_V^2\over 8}~~~~L_2^V,\gamma_2^{II}={g_V^2\over 4}
{}~~~L_3^V,\gamma_3^{II}=-{3\over 4}g_V^2,
\ee
which give the identities $L_2^V=2L_1^V$ and $L_3^V=-3L_2^V$.

Local term \rref{III.} fixes the vector contribution to the low energy
parameter $L_9$ (which also corresponds to the coefficient
$\gamma_9^{II}$ of the same local term in \rcite{EQUIV}) to be:
\be
L_9^V = {f_Vg_V\over 2}.
\ee
We thus derive the same relations as those obtained earlier.

\section{Conclusions}

In this letter we have shown explicitly the relation between a standard
vector field transforming as a vector and as a antisymmetric tensor
field.  We can thus immediately obtain the Lagrangians that are exactly
equivalent in both pictures. The relation of the vector representation
used here to the Hidden gauge model and others can be found
in\rcite{EQUIV,Yamawaki}.

The present work has added to the list of known field redefinitions also
the one that ends up with the tensor representation.  The method here
can be easily generalized to terms that contain powers of quark masses
and derivatives beyond those explicitly considered here, as well as to
the ''anomalous'' or abnormal intrinsic parity sector of vector meson
Lagrangians. The extension to axial vector mesons is similarly trivial.

\section*{Acknowledgments}

This work was partially supported by NorFA grant 93.15.078/00.  JB
thanks IPN-Orsay, where part of this work was done, for hospitality.
The work of EP was supported by the EU Contract Nr. ERBCHBGCT 930442.

\end{document}